# Reconciling γ-band Fermi-surface topologies in bilayer nickelate thin films: evidence for a three-dimensional Lifshitz transition

**Pengxu Ran[1]*, Jiexiong Sun[1]*, Runnan Wu[1]*, Haoyuan Yan[1], Zhen Wang[2], Tongrui Li[1], Xinzhe Luo[1], Mao Ye[3], Jishan Liu[3], Zhengtai Liu[3], Jing Tao[1,2], Donglai Feng[4†] and Juan Jiang[1,4†]**

[1]*Hefei National Research Center for Physical Sciences at Microscale and School of Emerging Technology, University of Science and Technology of China, Hefei, 230026, China*

[2]*Department of Physics, University of Science and Technology of China, Hefei, Anhui 230026, China*

[3]*Shanghai Synchrotron Radiation Facility, Shanghai Advanced Research Institute, Chinese Academy of Sciences, Shanghai 201210, China*

[4]*New Cornerstone Science Laboratory, Hefei National Laboratory, Hefei 230088, China*

[*] These authors contributed equally.

[†] E-mail: jjiangcindy@ustc.edu.cn, dlfeng@hfnl.cn

## Abstract

Despite extensive studies of bilayer nickelates, a unified picture of the Fermi-surface topology of the $d_{z^2}$-derived γ band remains lacking. Here, we use photon-energy-dependent angle-resolved photoemission spectroscopy to investigate $La_{1.5}Pr_{1.5}Ni_2O_7$ thin films across a series of epitaxial strain states. From tensile to compressive strain, the $d_{x^2-y^2}$-derived β Fermi surface systematically expands, while the flat γ-band top shifts to lower energy. With its pronounced three-dimensional dispersion, the γ band crosses the Fermi energy only over part of the measured $k_z$ range in the strongly compressively strained, superconducting $SrLaAlO_4$-grown film, but throughout the measured $k_z$ range under the other strain conditions, revealing a strain-driven three-dimensional Lifshitz transition. Oxygen deficiency similarly shifts the γ band upward, tuning it across the same Lifshitz boundary. Under both strain and oxygen tuning, γ-band changes are accompanied by pronounced differences in the Hall response, suggesting a connection to multiband transport. Together, these results support epitaxial strain and oxygen deficiency as distinct means of tuning three-dimensional γ-band Fermi-surface topology. More broadly, this strain-oxygen-$k_z$ framework helps reconcile disparate γ-band observations and guides further exploration of their connection to superconductivity.

## Introduction

Bilayer Ruddlesden–Popper (RP) nickelates provide a multiorbital platform in which lattice structure and low-energy electronic states can be tuned by external pressure and epitaxial strain. $La_3Ni_2O_7$ exhibits high-temperature superconductivity under pressure, opening new opportunities for investigating unconventional superconductivity[1-5]. More recently, ambient-pressure superconductivity has been realized in thin films of $La_3Ni_2O_7$ and rare-earth-substituted bilayer nickelates under epitaxial strain, with reported onset temperatures of 26–42 K in early studies and above 60 K in more recent reports[6-12]. Unlike bulk samples under high pressure, these thin films provide direct access to their low-energy electronic structure by angle-resolved photoemission spectroscopy (ARPES), while epitaxial strain offers a means of systematically tuning their electronic structure and Fermi surface.

Bilayer nickelates host multiple Ni-derived states near the Fermi level, with predominantly $d_{x^2-y^2}$-derived α and β bands and $d_{z^2}$-derived γ band. The bilayer structure enhances interlayer hybridization between Ni $d_{z^2}$ orbitals through the apical oxygen, producing a pronounced bonding–antibonding splitting and a relatively flat γ band near the Fermi energy ($E_F$)[13-15]. A previous photoemission study has further established appreciable $k_z$ dispersion of the γ band, highlighting its three-dimensional character[16]. Because the γ band lies close to $E_F$, changes in its energy and dispersion can substantially modify the associated Fermi-surface topology. Its possible involvement in superconductivity has motivated extensive theoretical studies[17-27], although pairing mechanisms that do not require a γ Fermi-surface pocket have also been proposed[22,28].

Experimentally, however, no unified picture has emerged for the γ-band Fermi-surface topology in bilayer nickelates. In bulk samples at ambient pressure, the flat γ-band top has been reported below $E_F$[13,14], whereas thin-film measurements have variously observed a γ band crossing $E_F$[15,29] or a flat-band top located 50 – 75 meV below $E_F$[30,31]. These apparently different observations may reflect variations in epitaxial strain, oxygen content and the probed $k_z$ position, but how these parameters reshape the three-dimensional γ Fermi surface remains unclear. This question is particularly relevant because ambient-pressure superconductivity in bilayer nickelate thin films has so far been predominantly associated with compressive epitaxial strain[6,11], and oxygen content provides an additional material parameter that can modify both the electronic structure and superconductivity[7,32,33]. A systematic determination of how strain and oxygen content tune the three-dimensional γ-band Fermi surface, and how these changes are reflected in charge transport, is therefore needed.

Here, we use photon-energy-dependent ARPES to systematically investigate $La_{1.5}Pr_{1.5}Ni_2O_7$ (LPNO) thin films across a range of epitaxial strains. By comparing the low-energy bands at different photon energies, we establish their distinct dimensionalities and track the strain evolution of the α, β and γ bands. The γ band exhibits pronounced $k_z$ dispersion, and its Fermi-level crossing evolves systematically with strain, consistent with a strain-driven Lifshitz transition of its three-dimensional Fermi surface. This evolution is accompanied by a sign change in the Hall coefficient, indicating a concomitant reconstruction of multiband transport. We further examine an oxygen-deficient $SrLaAlO_4$ (SLAO)-grown LPNO film. At the $k_z$ sections where the γ band lies below $E_F$ in the superconducting film, it crosses $E_F$ in the oxygen-deficient film, consistent with oxygen-driven tuning across the same Lifshitz boundary. The accompanying change in the Hall response shows that this reconstruction coincides with a change in multiband charge transport. Together, these findings establish epitaxial strain and oxygen content as distinct parameters for tuning the multiband electronic structure of bilayer nickelates.

## Growth and characterization of LPNO thin films

Three-unit-cell-thick (3-UC) LPNO films were grown on $SrTiO_3$ (STO), $(LaAlO_3)_{0.3}(Sr_2AlTaO_6)_{0.7}$ (LSAT), $LaAlO_3$ (LAO), and SLAO substrates by oxygen-plasma-assisted oxide molecular-beam epitaxy (OMBE), as illustrated in Fig. 1a. The growth process was monitored in situ by reflection high-energy electron diffraction (RHEED), with sharp and well-defined post-growth patterns indicating smooth and well-ordered film surfaces (details in Fig. S1). High-resolution X-ray diffraction (HRXRD) measurements in Fig. 1b show well-defined 00l reflections characteristic of the n = 2 RP phase, with no additional reflections associated with other RP phases. The out-of-plane lattice constants extracted from the film diffraction peaks are summarized in Fig. 1c and are comparable to those previously reported for strain-engineered bilayer nickelate thin films[11]. Reciprocal-space maps (RSMs) collected around selected asymmetric reflections in Fig. 1d show that the in-plane reciprocal-space coordinates of the films coincide with those of their respective substrates, demonstrating that the 3-UC films remain fully strained. The corresponding nominal in-plane misfit strains relative to bulk $La_3Ni_2O_7$ are +1.9%, +0.9%, −1.2%, and −2.0% for STO, LSAT, LAO, and SLAO, respectively[6,11].

The temperature-dependent resistance curves of the LPNO films, normalized to their values at 300 K, are presented in Fig. 1e. All samples were annealed under oxygen plasma after growth to minimize the oxygen deficiency. A superconducting transition with an onset temperature of approximately 30 K is observed only in the SLAO-grown film, whereas no superconducting transition is detected under the other strain conditions. Atomic-resolution high-angle annular dark-field scanning transmission electron microscopy (HAADF-STEM) of the SLAO-grown film in Fig. 1f further resolves the repeated bilayer stacking of $NiO_2$ layers separated by rock-salt layers, confirming the bilayer–bilayer (2222) structure within the imaged region. Together, these structural and transport measurements establish a well-characterized strain series in which the epitaxial strain is systematically varied while the films retain the targeted bilayer RP structure, providing a controlled platform for investigating the evolution of their low-energy electronic structure.

## Electronic structure of the STO-grown LPNO film

We first characterize the three-dimensional electronic structure of the tensile-strained STO-grown LPNO film using photon-energy-dependent ARPES. To minimize oxygen loss, all the samples were annealed in oxygen plasma and then transported to the synchrotron facility in a cryogenic ultrahigh-vacuum transfer suitcase for ARPES measurements (see Methods). Figure 2a shows an out-of-plane Fermi surface map along the in-plane $\bar{\mathrm{M}}-\bar{\mathrm{X}}$ direction, acquired by varying the photon energy from 49 to 180 eV. The β band remains visible over most of the photon-energy range, with weak $k_z$ dependence indicating its quasi-two-dimensional character. In contrast, the γ-band intensity exhibits a pronounced periodic modulation along $k_z$, with strong suppression and enhancement at different photon energies. The observed modulation period of approximately 1.645 $Å^{-1}$ corresponds to an effective real-space length scale of about 3.82 Å, comparable to the intra-bilayer Ni–Ni spacing. This modulation is a matrix-element effect associated with the bilayer bonding state, arising from interference between photoemission amplitudes from the two $NiO_2$ layers [30,34,35].

The strong matrix-element modulation provides distinct photon-energy windows in which the β and γ bands can be selectively resolved. At photon energies near 103 eV, the γ-band intensity is strongly suppressed and the spectral weight near $E_F$ is dominated by the β band. By contrast, the γ band is strongly enhanced near photon energies of 75 and 153 eV, corresponding to distinct $k_z$ sections at different phases of the intensity modulation. We therefore use a photon energy of 103 eV to resolve the β band and photon energies of 75 and 153 eV to examine the γ band.

Figures 2b–d and 2f–h show the corresponding Fermi surface maps and band dispersions along $\bar{\mathrm{M}} - \bar{\mathrm{X}}$. The β band is clearly resolved at a photon energy of 103 eV and a weaker β-band signal remains discernible in the momentum distribution curves (MDCs) at a photon energy of 75 eV. In contrast, the γ-band dispersion and Fermi crossing are clearly resolved at both photon energies, 75 and 153 eV. The pronounced difference in the γ-band Fermi-crossing momenta at these two photon energies demonstrates the intrinsic $k_z$ dependence of the γ band, beyond the photon-energy-dependent matrix-element modulation. Importantly, the γ band crosses $E_F$ at both $k_z$ sections, indicating that the γ Fermi surface extends across the measured $k_z$ range in the tensile-strained STO-grown film. The α Fermi-surface pocket is also resolved at a photon energy of 75 eV in Fig. 2c with a square-like Fermi-surface shape, and the α-band dispersion can be clearly identified in the corresponding cut along $\bar{\Gamma} - \bar{\mathrm{M}}$ in Fig. 2e. Together, these measurements establish the low-energy electronic structure of the STO-grown film: the $d_{x^2-y^2}$-derived β band is predominantly quasi-two-dimensional, whereas the $d_{z^2}$-derived γ band exhibits pronounced $k_z$ dispersion while maintaining a Fermi-level crossing throughout the measured $k_z$ range. An additional nearly vertical spectral-weight feature, labeled γ′, is observed near $\bar{\mathrm{M}}$ at a photon energy of 153 eV and may arise from electronic correlations[29-31].

**Strain tuning of the electronic structure in LPNO thin films**

We next investigate how epitaxial strain reconstructs the low-energy electronic structure of LPNO films. Similar γ-band intensity modulation is also observed in the compressively strained LAO-grown film (Fig. S2), indicating that this matrix-element effect is intrinsic to the bilayer electronic structure rather than specific to strain. Figures 3a-d show the Fermi surface maps and corresponding band dispersions along $\bar{\mathrm{M}} - \bar{\mathrm{X}}$ measured at selected photon energies for films grown on different substrates. The Fermi surface maps and corresponding band dispersions in Figs. 3a and 3b were measured at a photon energy of 103 eV, where the β-band intensity dominates. As strain evolves from tensile to compressive, the β pocket systematically expands, as shown by the fitted MDCs near $E_F$ in Fig. 3e. Correspondingly, the separation (δk) between the two β-band

Fermi crossings on either side of $\bar{X}$ decreases from tensile to compressive strain. This trend is consistent with the β-band dispersions calculated using density functional theory (DFT) (Figs. S3 and S4), demonstrating a systematic evolution of the β-pocket size with epitaxial strain. The β-pocket sizes of the STO- and SLAO-grown films show finite photon-energy dependence (Fig. S5), indicating that the β band is not strictly two-dimensional. Nevertheless, its comparatively weak $k_z$ dispersion supports a predominantly quasi-two-dimensional description under both tensile and compressive strain.

The γ band undergoes a pronounced evolution under strain. Figures 3c and 3d show Fermi surface maps and band dispersions measured at a photon energy of 153 eV, except for the SLAO map acquired at a photon energy of 160 eV to better resolve the β pocket. Clear γ-band Fermi crossings are resolved for the STO-, LSAT-, and LAO-grown films at a photon energy of 153 eV, as shown by the ARPES spectra and the corresponding MDCs near $E_F$ (Figs. 2h and 3d). In contrast, no distinct γ-band Fermi-crossing peaks are resolved for the strongly compressively strained SLAO-grown film at this photon energy (Fig. 3d). The γ-band top in the SLAO-grown film is estimated to lie approximately 50 meV below $E_F$ from the background-subtracted energy distribution curve (EDC) (details in Fig. S6), comparable to the values reported previously[30,31]. Moreover, the fitted γ-band dispersion slope in Fig. 3f decreases systematically with increasing compressive strain, indicating that the γ-band dispersion becomes progressively flatter as it shifts to lower energy (details in Fig. S7). These observations indicate that compressive strain progressively shifts the γ-band energy downward, consistent with the trend obtained from DFT calculations (Fig. S3).

The absence of a γ-band crossing in the SLAO-grown film at a photon energy of 153 eV, however, does not imply that the γ band is entirely below $E_F$. To determine its $k_z$-dependent Fermi-level crossing, we further examine the same SLAO-grown film at a photon energy of 75 eV (Figs. 3g, h), corresponding to a different $k_z$ section. At this photon energy, the β and γ bands are clearly separated in momentum, and the MDC near $E_F$ in Fig. 3g reveals a distinct γ-band Fermi crossing. Thus, in the strongly compressively strained SLAO-grown film, the γ band crosses $E_F$ at some $k_z$ values but remains below $E_F$ at others. In contrast, the γ band crosses $E_F$ at all the measured $k_z$ sections in the STO-, LSAT-, and LAO-grown films. The continuous downward shift of the $k_z$-dispersive γ band with increasing compressive strain therefore indicates evolution toward a three-dimensional Lifshitz transition, with the superconducting

SLAO-grown film residing in a near-Lifshitz regime where the γ band lies close to $E_F$ and its Fermi surface extends over only part of the measured $k_z$ range.

## Transport responses accompanying strain- and oxygen-tuned electronic structures

The strain-dependent measurements above show that the superconducting SLAO-grown film has a three-dimensional Fermi surface close to the γ-band Lifshitz transition. Because the electronic properties of SLAO-grown LPNO films are highly sensitive to oxygen content[7,32,33], we next examine how oxygen deficiency affects this near-Lifshitz electronic structure in films grown on the same substrate. Figure 4a presents the normalized temperature-dependent resistance of an oxygen-deficient SLAO-grown LPNO film. In contrast to the superconducting SLAO-grown film, the oxygen-deficient film shows no superconducting transition and exhibits a metal-to-insulator crossover upon cooling, consistent with previous reports[32,36]. The corresponding band dispersions along $\bar{\mathrm{M}}-\bar{\mathrm{X}}$ measured at photon energies of 103 and 153 eV are shown in Fig. 4b, predominantly resolving the β and γ bands. The γ-band electronic structure is markedly distinct from that of the superconducting film. Figure 4c compares the MDCs near $E_F$ of the two films. At a photon energy of 103 eV, the β-band Fermi wave vector varies only slightly between the two films. In contrast, at a photon energy of 153 eV, clear γ-band Fermi-crossing peaks are observed in the oxygen-deficient film, whereas the γ-band top lies below $E_F$ in the superconducting film. This comparison is consistent with oxygen-driven tuning across the three-dimensional Lifshitz boundary identified from the strain series.

The Hall response is summarized in Fig. 4d, together with the strain-dependent Hall coefficients (see Fig. S8 for details). The superconducting SLAO-grown film exhibits a negative Hall coefficient, whereas both the oxygen-deficient SLAO-grown film and LAO-, LSAT-, and STO-grown films show a positive Hall coefficient. The Hall response therefore changes sign not only across the strain series but also upon oxygen tuning at fixed epitaxial strain. The concomitant evolution of the γ-band electronic structure and Hall response suggests a connection between γ-band reconstruction and multiband charge transport.

## Discussion

Figure 4e summarizes the unified evolution of the three-dimensional γ-band Fermi surface under epitaxial-strain and oxygen-content tuning. Starting from the superconducting SLAO-grown film, which lies close to the γ-band Lifshitz boundary, both reduced compressive strain and oxygen deficiency shift the γ band upward. Reducing compressive strain extends the Fermi-level crossings from a restricted set of $k_z$ sections to all measured sections, whereas the γ band in the oxygen-deficient SLAO-grown film crosses $E_F$ at a $k_z$ section where it lies below $E_F$ in the superconducting film. Thus, two distinct material parameters can tune the γ band across the same three-dimensional Lifshitz boundary. More broadly, this framework can reconcile disparate γ-band observations, although the full strain – oxygen – $k_z$ parameter space remains to be mapped.

Epitaxial strain modifies the low-energy electronic structure of LPNO films, as reflected in the evolution of the β and γ bands. Previous studies show that strain-induced changes in the $NiO_6$ octahedral geometry can modify the local crystal-field environment and orbital hopping[11,37], providing a possible explanation for the observed changes in the band dispersions. In particular, the $k_z$-dispersive γ band shifts downward with increasing compressive strain, consistent with a three-dimensional Lifshitz transition as its band top passes through $E_F$ at selected $k_z$ sections. Oxygen tuning provides a distinct route to the same γ-band reconstruction. In SLAO-grown films, the γ band lies higher in the oxygen-deficient film than in the superconducting film, consistent with the strong sensitivity of the $d_{z^2}$-derived state to the out-of-plane hopping through the apical oxygen. Oxygen vacancies, particularly at inner-apical sites, may modify the local Ni–O coordination and interlayer hopping, providing a possible mechanism for changes in the γ-band electronic structure[38-40]. The Hall coefficient changes sign concurrently with the γ-band reconstruction under both strain and oxygen tuning, suggesting a connection between the γ-band evolution and multiband charge transport, although the Hall response reflects the combined contributions of multiple Fermi-surface sheets and their scattering and mobility.

The superconducting SLAO-grown film therefore resides in a particularly interesting Fermiological regime, where the $d_{z^2}$-derived γ band lies close to $E_F$ and its three-dimensional Fermi surface is near a Lifshitz transition. Recent theoretical work has suggested that the γ pocket can introduce competing pair-forming and pair-breaking scattering channels, with enhanced pairing when the $d_{z^2}$-derived band lies close to $E_F$ near the Lifshitz regime[21]. Conversely, tensile strain and a more extended γ Fermi surface may enhance nesting-related scattering channels between α and γ, which have been proposed to suppress superconducting pairing[41]. Our observations therefore suggest that the energy position and three-dimensional topology of the γ band may be relevant to the superconducting state, while determining the optimal γ-band configuration for superconductivity requires further theoretical and experimental investigation.

## Methods

### Thin-film growth and annealing

LPNO films were grown on (001)-oriented STO, LSAT, LAO, and SLAO substrates by oxygen-plasma-assisted OMBE. La, Pr, and Ni were supplied from independent effusion sources. A quartz crystal microbalance (QCM) was used to monitor the source fluxes. The fluxes were calibrated using binary-oxide reference films[42,43]. La and Pr oxides were grown on Y-$ZrO_2$(111), whereas NiO was grown on MgO(100). The cation fluxes were determined from the film thicknesses measured by X-ray reflectivity, and the La:Pr flux ratio was fixed at 1:1.

Following established shuttered-growth strategies that account for layer rearrangement and interface stoichiometry in Ruddlesden–Popper oxides[44,45], we employed a repeated [(La,Pr)O–(La,Pr)O–$NiO_2$–$NiO_2$–(La,Pr)O] shutter sequence. The films were grown at 720 °C under an oxygen pressure of $5\times10^{-6}$ mbar with an oxygen-plasma power of 150 W. The growth process was monitored in situ by RHEED.

After growth, the films were annealed in the same chamber for 1 h at 300 °C under an oxygen pressure of $3\times10^{-5}$ mbar with an oxygen-plasma power of 300 W. For the oxygen-deficient sample, the annealing oxygen pressure and plasma power were reduced to $5\times10^{-6}$ mbar and 150 W, respectively, while the annealing temperature and duration remained unchanged. Following rapid cooling, the samples were loaded into a cryogenic ultrahigh-vacuum transfer suitcase within 15 min and maintained at approximately 120 K during transport to minimize oxygen loss. The suitcase had a base pressure below $3\times10^{-10}$ mbar.

### Structural, microscopic, and transport characterization

HRXRD measurements were conducted using a Rigaku SmartLab diffractometer equipped with a 9-kW Cu Kα X-ray source. The crystalline phase and out-of-plane orientation were examined by θ–2θ scans, and the out-of-plane lattice constants were determined from the film diffraction peak positions. RSMs around selected asymmetric reflections were collected after azimuthal alignment to determine the epitaxial strain states of the films.

Cross-sectional STEM specimens approximately 60 nm thick were prepared using a Thermo Fisher Helios G5x focused ion beam system. Atomic-resolution HAADF-STEM imaging was performed on a double-spherical-aberration-corrected JEOL Grand ARM300 microscope operated at 300 kV, using a convergence semi-angle of 25 mrad and a collection semi-angle range of 41–200 mrad.

Electrical transport measurements were performed using a Quantum Design Physical Property Measurement System (PPMS). Temperature-dependent resistance was measured immediately after the ARPES measurements in a standard four-probe configuration with an excitation current of 5 μA. For Hall measurements, electrodes were deposited by evaporation in a Hall-bar geometry, and the transverse voltage was recorded with the magnetic field applied perpendicular to the film plane. The Hall resistance was obtained by anti-symmetrizing the transverse resistance with respect to the magnetic field.

**ARPES measurements**

ARPES measurements were performed at the BL03U beamline of Shanghai Synchrotron Radiation Facility (SSRF)[46]. The samples were transported using the cryogenic ultrahigh-vacuum transfer suitcase and loaded directly into the ARPES endstation without exposure to air. Photoelectrons were analyzed using a Scienta Omicron DA30 hemispherical electron analyzer. The measurement chamber was maintained at a base pressure below $7\times10^{-11}$ mbar. Measurements were conducted at 10 K using linearly polarized light in the horizontal geometry over a photon-energy range of 49–180 eV. The total energy resolution ranged from 10 to 20 meV, depending on the photon energy, and the angular resolution was better than 0.2°.

**DFT calculations**

Density functional theory calculations were performed using the projector augmented-wave method[47] implemented in the Vienna *ab initio* Simulation Package[48]. Exchange-correlation effects were treated using the generalized-gradient approximation in the Perdew-Burke-Ernzerhof[49] functional. A plane-wave cutoff energy of 450 eV was employed, and the Brillouin zone was sampled using a Monkhorst-Pack k-point grid of 12×12×12. The lattice constants were fixed to the experimentally determined values, while the internal atomic coordinates were fully relaxed.

**Acknowledgements**

This work is supported by the National Natural Science Foundation of China (Grant No. 12574071， Grant No. 12688301), the National Key R&D Program of China (Grant No. 2023YFA1406304), the HFNL Self-Deployed Project (Grant No. ZB2602000301), and the New Cornerstone Science Foundation. Part of this research used Beamline 03U of the Shanghai Synchrotron Radiation Facility, which is supported by the ME2 project under contract no. 11227902 from the National Natural Science Foundation of China.

**Author contributions**

D.F. and J.J. supervised the project. J.J. and P.R. conceived the project. P.R., J.S. and R.W. performed the ARPES measurements and analyzed the data, with assistance from M.Y., Z.L. and J.L. during the measurements. J.S., R.W., P.R. and H.Y. grew the thin films. J.S. and R.W. performed the transport measurements and analyzed the data. X.L. performed the X-ray diffraction measurements. Z.W. and J.T. performed the transmission electron microscopy measurements. T.L. performed the density functional theory calculations. P.R., D.F. and J.J. wrote the manuscript with input from all authors.

**Competing Interests**

The authors declare no competing interests.

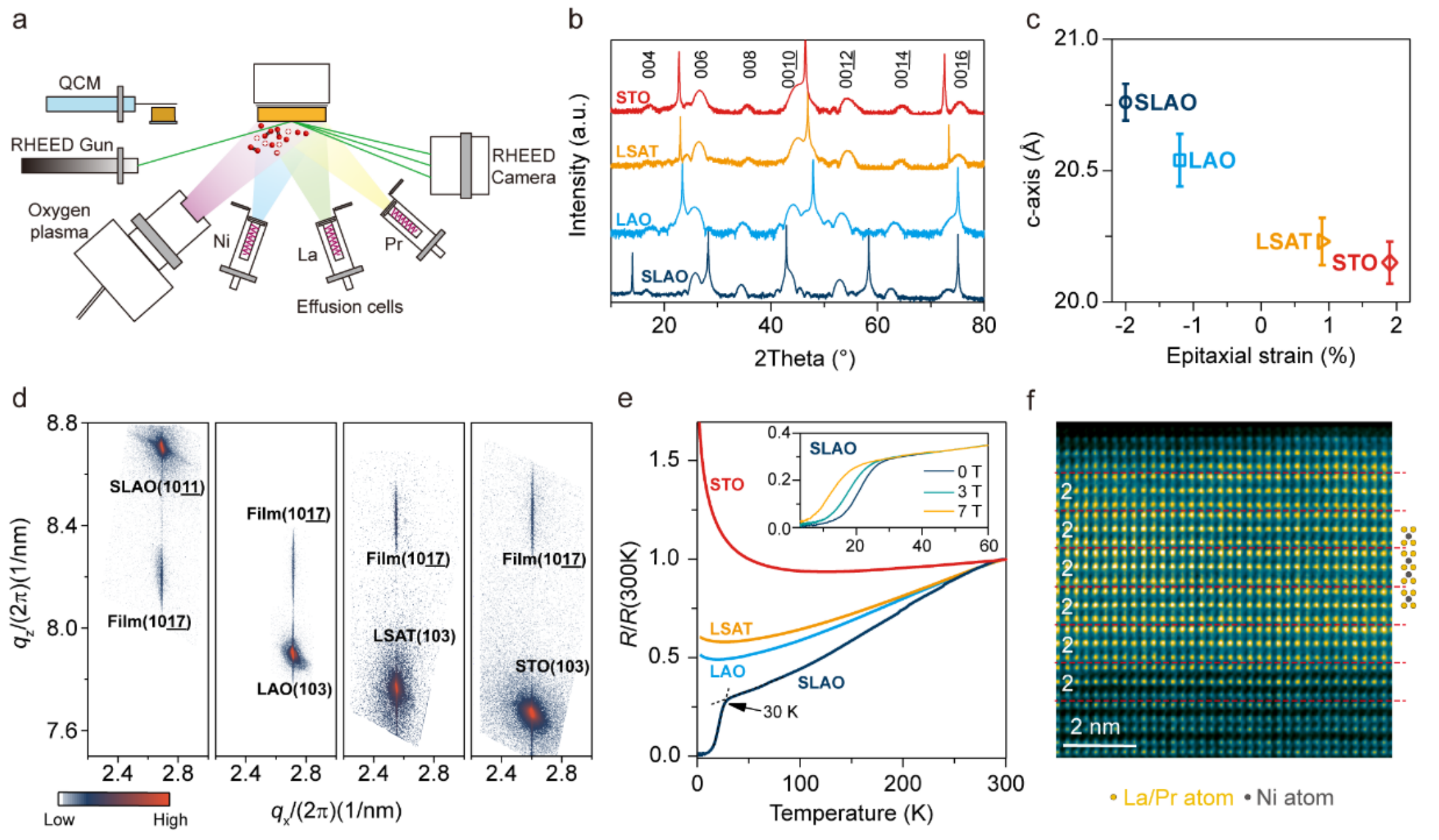

**Fig. 1 | Growth, structure, and transport properties of LPNO films under different epitaxial strains. a,** Schematic of the oxygen-plasma-assisted OMBE system used to grow three-unit-cell-thick LPNO films. **b,** HRXRD θ–2θ scans of films under different epitaxial strains. **c,** Out-of-plane lattice constants extracted from the film diffraction peaks in panel b, plotted as a function of epitaxial strain. **d,** Reciprocal-space maps around selected asymmetric reflections, showing that all films are fully strained to their respective substrates. **e,** Normalized resistance–temperature curves of films under different epitaxial strains. Only the SLAO-grown film exhibits a superconducting onset at approximately 30 K. Inset: suppression of the transition with increasing out-of-plane magnetic field. **f,** Cross-sectional atomic-resolution HAADF-STEM image of the SLAO-grown film along the [110] zone axis of the substrate, showing the bilayer–bilayer (2222) stacking sequence.

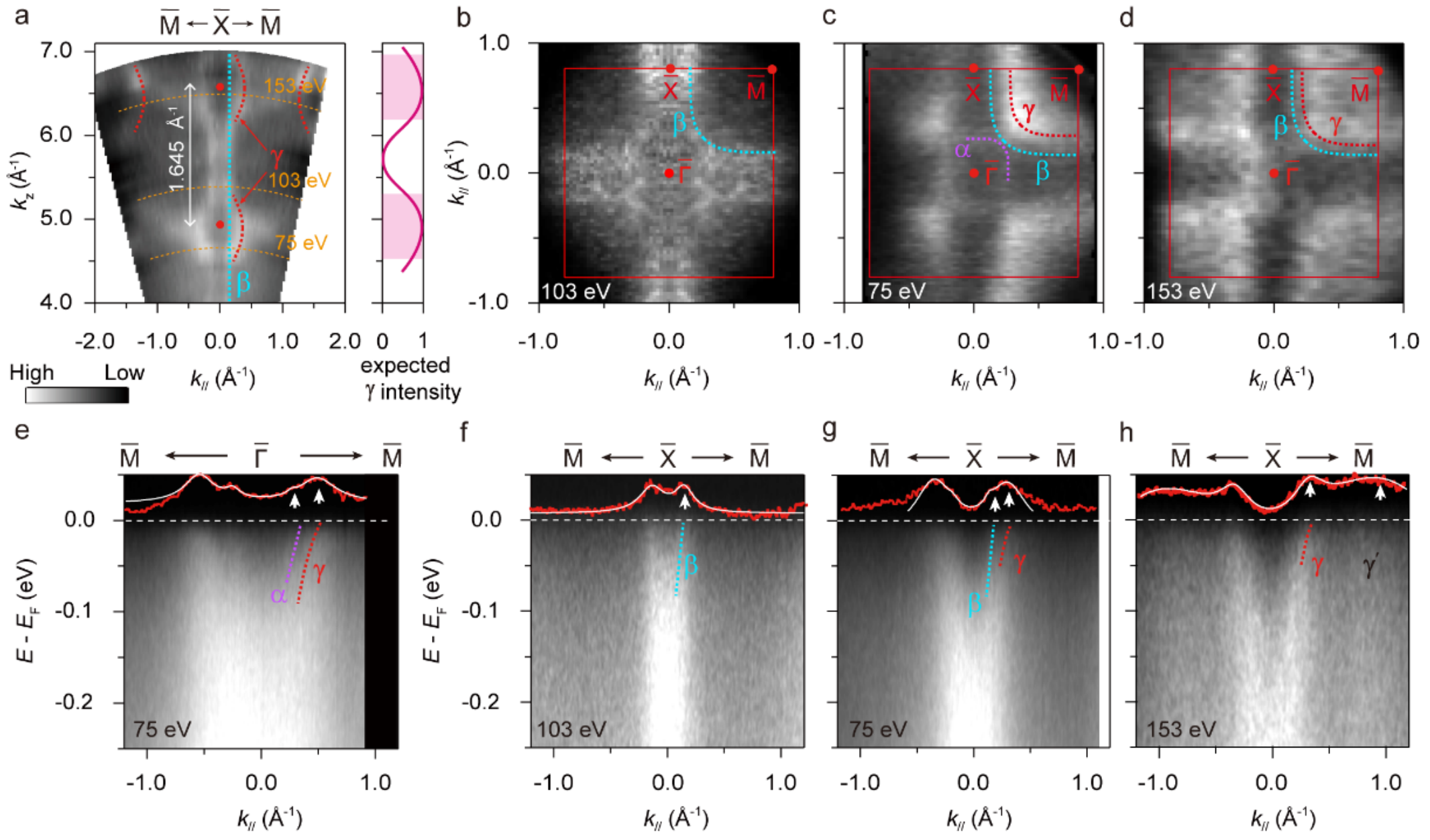

**Fig. 2 | Three-dimensional electronic structure of the tensile-strained STO-grown LPNO film. a,** Out-of-plane Fermi surface map along the in-plane $\overline{\mathrm{M}}-\overline{\mathrm{X}}$ direction, acquired by varying the photon energy from 49 to 180 eV. The $k_z$ values are estimated assuming an inner potential of 10 eV. The white double-headed arrow indicates the period of the γ-band spectral-intensity modulation. Orange dashed lines mark the measurement trajectories at photon energies of 75, 103 and 153 eV. The right panel shows the expected oscillation of the γ-band spectral intensity, with the pink shaded regions indicating where the γ-band signal is visible. **b–d,** Fermi surface maps acquired at photon energies of 103, 75 and 153 eV, respectively, using a 30-meV energy integration window. The red solid lines indicate the Brillouin-zone boundaries. **e,** ARPES spectrum along $\overline{\Gamma}-\overline{\mathrm{M}}$, acquired at a photon energy of 75 eV, showing the α and γ bands. **f–h,** ARPES spectra along $\overline{\mathrm{M}}-\overline{\mathrm{X}}$, acquired at photon energies of 103, 75 and 153 eV, respectively. A nearly vertical spectral feature, labeled γ′, is observed near $\overline{\mathrm{M}}$ in **h**. The red curves in panels e–h are MDCs extracted near $E_F$, with the corresponding fits shown as white solid lines. White arrows mark the corresponding MDC peak positions. The purple, cyan, and red dashed lines are guides to the eye for the α, β and γ bands, respectively.

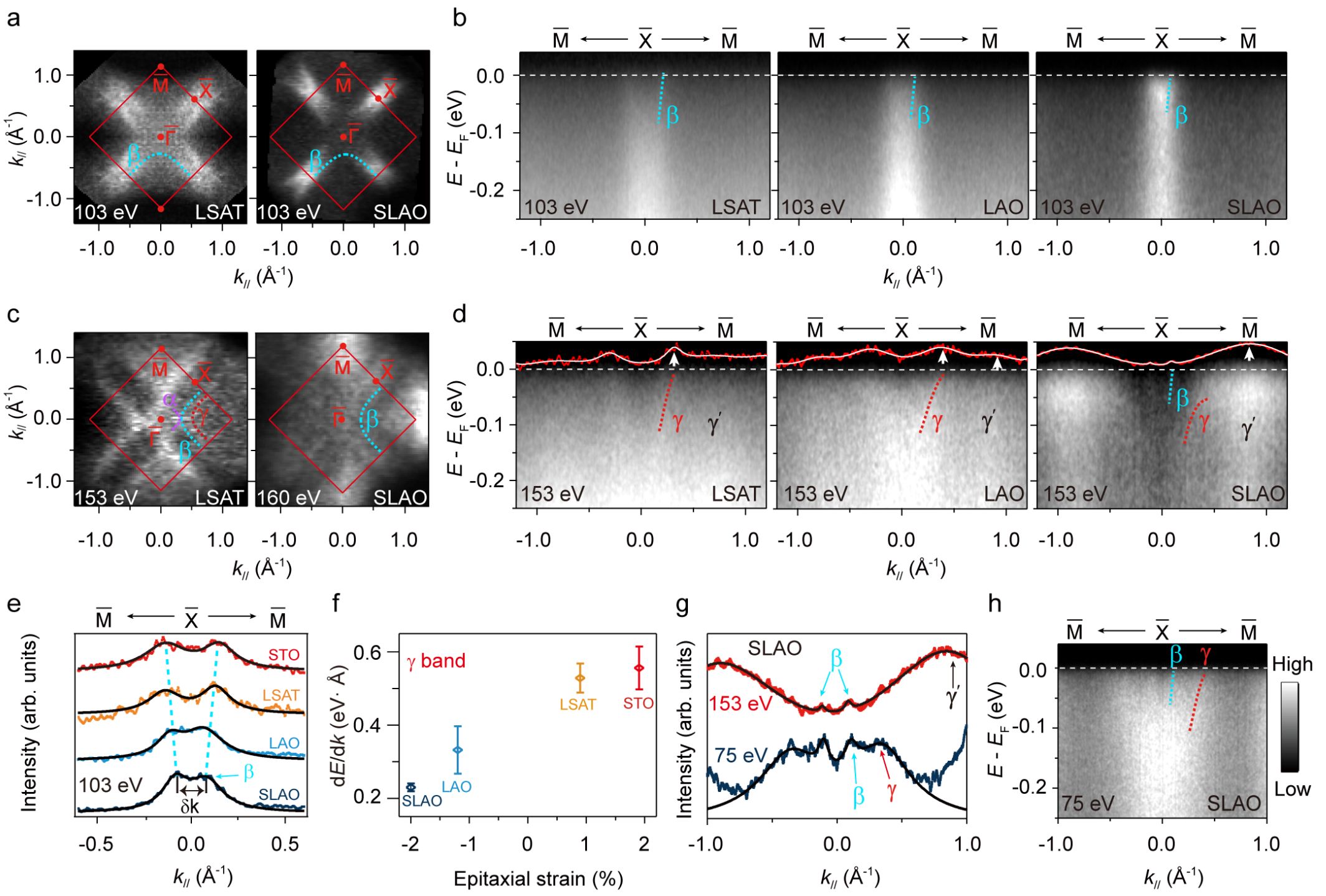

**Fig. 3 | Epitaxial-strain tuning of the electronic structure of LPNO thin films. a, b,** Fermi surface maps and ARPES cuts along $\bar{M}-\bar{X}$, respectively, for LPNO films under different epitaxial strains, measured at a photon energy of 103 eV. **c, d,** Corresponding Fermi surface maps and ARPES cuts measured at 153 eV, except for the SLAO-grown film map in **c**, which was acquired at 160 eV to better resolve the β pocket. MDCs near $E_F$ are shown in **d**, with the corresponding fits represented by white solid lines. The ARPES cut of the SLAO-grown film in panel **d** is background-subtracted (see **Fig. S6** for details). **e,** MDCs near $E_F$ from ARPES cuts along $\bar{M}-\bar{X}$ measured at a photon energy of 103 eV for films under different epitaxial strains. **f,** Average γ-band dispersion slopes along $\bar{M}-\bar{X}$ for films under different epitaxial strains, obtained from linear fits to the MDC-derived dispersions measured at a photon energy of 153 eV (see **Fig. S7** for details). **g,** Comparison of MDCs near $E_F$ for the SLAO-grown film measured at photon energies of 75 and 153 eV. **h,** ARPES cut of the SLAO-grown film along $\bar{M}-\bar{X}$ measured at a photon energy of 75 eV. A 30-meV energy integration window was used in **a** and **c**, and red solid lines indicate the Brillouin-zone boundaries. Dashed lines overlaid on the Fermi surface maps and ARPES cuts serve as guides to the eye for the band features. Black curves in **e** and **g** show the corresponding fits.

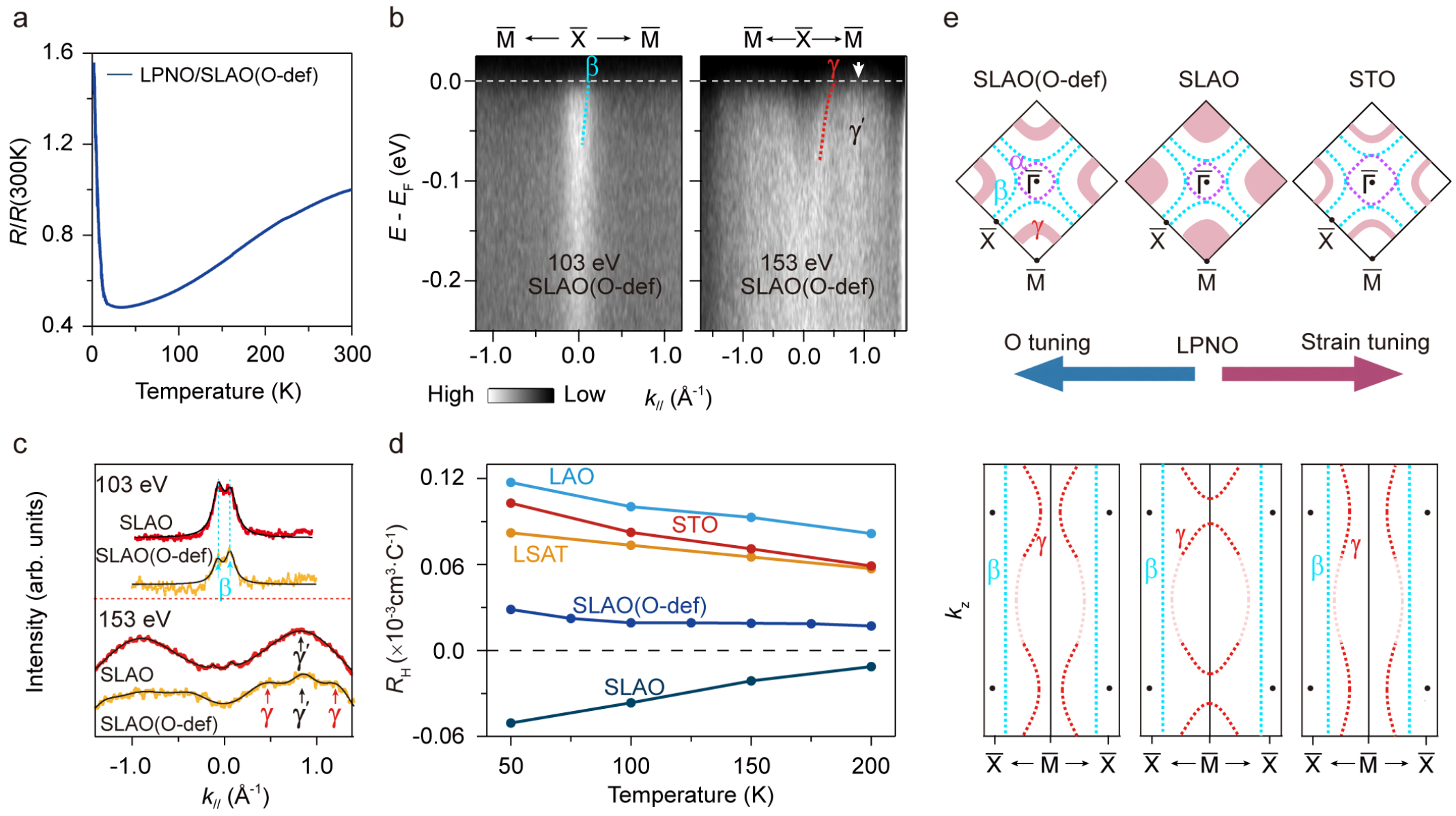

**Fig. 4 | Effects of epitaxial strain and oxygen content on the electronic structure and transport of LPNO films. a,** Temperature-dependent resistance of the oxygen-deficient SLAO-grown film (O-def), normalized to its value at 300 K. **b,** ARPES cuts of the oxygen-deficient SLAO-grown film along $\bar{\mathrm{M}}-\bar{\mathrm{X}}$, measured at photon energies of 103 and 153 eV. Cyan and red dashed lines serve as guides to the eye for the β and γ bands, respectively. **c,** Comparison of MDCs near $E_F$ for the superconducting and oxygen-deficient SLAO-grown films at photon energies of 103 and 153 eV. Black curves show the corresponding fits. **d,** Temperature dependence of the Hall coefficients of LPNO films under different epitaxial strains, including the superconducting and oxygen-deficient films grown on SLAO. **e,** Schematic illustration of the Fermi-surface evolution under epitaxial-strain and oxygen-content tuning. The upper and lower panels illustrate the in-plane and out-of-plane Fermi surfaces, respectively. In the upper panels, purple and cyan dashed lines serve as guides to the eye for the α- and β-band contours, respectively, and the shaded regions indicate the γ-band regions. In the lower panels, cyan and red dashed lines serve as guides to the eye for the β and γ bands, respectively. The darker segments of the red dashed lines indicate regions where the γ-band spectral intensity is experimentally observed.

## References

1 Sun, H. *et al.* Signatures of superconductivity near 80 K in a nickelate under high pressure. *Nature* **621**, 493-498 (2023).

2 Hou, J. *et al.* Emergence of High-Temperature Superconducting Phase in Pressurized $La_3Ni_2O_7$ Crystals. *Chin. Phys. Lett.* **40**, 117302 (2023).

3 Zhang, Y. *et al.* High-temperature superconductivity with zero resistance and strange-metal behaviour in $La_3Ni_2O_{7-\delta}$. *Nat. Phys.* **20**, 1269-1273 (2024).

4 Wang, G. *et al.* Pressure-Induced Superconductivity in Polycrystalline $La_3Ni_2O_{7-\delta}$. *Phys. Rev. X* **14**, 011040 (2024).

5 Wang, N. *et al.* Bulk high-temperature superconductivity in pressurized tetragonal $La_2PrNi_2O_7$. *Nature* **634**, 579-584 (2024).

6 Ko, E. K. *et al.* Signatures of ambient pressure superconductivity in thin film $La_3Ni_2O_7$. *Nature* **638**, 935-940 (2025).

7 Zhou, G. *et al.* Ambient-pressure superconductivity onset above 40 K in $(La,Pr)_3Ni_2O_7$ films. *Nature* **640**, 641-646 (2025).

8 Li, Q. *et al.* Enhanced superconductivity in the compressively strained bilayer nickelate thin films by pressure. *Nat. Commun.* **17**, 3276 (2026).

9 Wang, Z. *et al.* Electron-like high-temperature superconductivity induced by compressive strain in $La_2PrNi_2O_7$ thin films. *arXiv:2608.01295* (2026).

10 Han, Z., Xiang, L., Zhou, X. & Zhu, Z. Granular Superconductivity in $La_2PrNi_2O_{7-\delta}$ Thin Films. *Chin. Phys. Lett.* **43**, 080706 (2026).

11 Bhatt, L. *et al.* Structural modifications in strain-engineered bilayer nickelate thin films. *Nature* **653**, 76-82 (2026).

12 Zhou, G. *et al.* Superconductivity onset above 60 K in ambient-pressure nickelate films. *Natl. Sci. Rev.* **13**, nwag151 (2026).

13 Yang, J. *et al.* Orbital-dependent electron correlation in double-layer nickelate $La_3Ni_2O_7$. *Nat. Commun.* **15**, 4373 (2024).

14 Li, Y. *et al.* Electronic Correlation and Pseudogap-Like Behavior of High-Temperature Superconductor $La_3Ni_2O_7$. *Chin. Phys. Lett.* **41**, 087402 (2024).

15 Li, P. *et al.* Angle-resolved photoemission spectroscopy of superconducting $(La,Pr)_3Ni_2O_7/SrLaAlO_4$ heterostructures. *Natl. Sci. Rev.* **12**, nwaf205 (2025).

16 Li, Y. *et al.* Three-Dimensional Electronic Structures in Superconducting Ruddlesden-Popper Bilayer Nickelate Films. *Phys. Rev. X* **16**, 031016 (2026).

17 Luo, Z., Hu, X., Wang, M., Wu, W. & Yao, D. X. Bilayer Two-Orbital Model of $La_3Ni_2O_7$ under Pressure. *Phys. Rev. Lett.* **131**, 126001 (2023).

18 Zhang, Y., Lin, L. F., Moreo, A., Maier, T. A. & Dagotto, E. Structural phase transition, s±-wave pairing, and magnetic stripe order in bilayered superconductor $La_3Ni_2O_7$ under pressure. *Nat. Commun.* **15**, 2470 (2024).

19 Yang, Q.-G., Wang, D. & Wang, Q.-H. Possible s±-wave superconductivity in $La_3Ni_2O_7$. *Phys. Rev. B* **108**, L140505 (2023).

20 Lechermann, F., Gondolf, J., Bötzel, S. & Eremin, I. M. Electronic correlations and superconducting instability in $La_3Ni_2O_7$ under high pressure. *Phys. Rev. B* **108**, L201121 (2023).

21 Ryee, S., Witt, N., Sangiovanni, G. & Wehling, T. O. Superconductivity Governed by Janus-Faced Fermiology in Strained Bilayer Nickelates. *Phys. Rev. Lett.* **135**, 236003 (2025).

22 Gao, Y. Robust s±-wave pairing in a bilayer two-orbital model of pressurized $La_3Ni_2O_7$ without the γ Fermi surface. *Physica C* **640**, 1354824 (2026).

23 Qin, Q. & Yang, Y.-F. High-Tc superconductivity by mobilizing local spin singlets and possible route to higher Tc in pressurized $La_3Ni_2O_7$. *Phys. Rev. B* **108**, L140504 (2023).

24 Yang, Y.-F., Zhang, G.-M. & Zhang, F.-C. Interlayer valence bonds and two-component theory for high-Tc superconductivity of $La_3Ni_2O_7$ under pressure. *Phys. Rev. B* **108**, L201108 (2023).

25 Jiang, R., Hou, J., Fan, Z., Lang, Z. J. & Ku, W. Pressure Driven Fractionalization of Ionic Spins Results in Cupratelike High-Tc Superconductivity in $La_3Ni_2O_7$. *Phys. Rev. Lett.* **132**, 126503 (2024).

26 Lu, C., Pan, Z., Yang, F. & Wu, C. Interlayer-Coupling-Driven High-Temperature Superconductivity in $La_3Ni_2O_7$ under Pressure. *Phys. Rev. Lett.* **132**, 146002 (2024).

27 Heier, G., Park, K. & Savrasov, S. Y. Competing dxy and s± pairing symmetries in superconducting $La_3Ni_2O_7$: LDA+FLEX calculations. *Phys. Rev. B* **109**, 104508 (2024).

28 Shao, Z.-Y. *et al.* Pairing without γ-pocket in the $La_3Ni_2O_7$ thin film. *Commun. Phys.* (2026).

29 Nie, Z. *et al.* Superconductivity and electronic structures of nickelate thin film superstructures. *Nature* **652**, 628-634 (2026).

30 Wang, B. Y. *et al.* Electronic Structure of Compressively Strained Bilayer Nickelate Thin Film. *Phys. Rev. X* **16**, 031008 (2026).

31 Sun, W. *et al.* Observation of superconductivity-induced leading-edge gap in a bilayer nickelate. *Nat. Phys.* **22**, 1409-1415 (2026).

32 Miao, Y. *et al.* Electronic structures across superconductor-insulator transition in Ruddlesden-Popper bilayer nickelate films. *arXiv:2502.18068* (2026).

33 Zhang, S. *et al.* Interlayer hybridization promotes superconductivity in bilayer nickelates. *Nat. Mater.* (2026).

34 Moser, S. An experimentalist's guide to the matrix element in angle resolved photoemission. *J. Electron Spectrosc. Relat. Phenom.* **214**, 29-52 (2017).

35 Moreschini, L. *et al.* Bilayer splitting and wave functions symmetry in $Sr_3Ir_2O_7$. *Phys. Rev. B* **89**, 201114 (2014).

36 Wang, M. *et al.* Superconducting Dome in La3-xSrxNi$_2$O$_{7-\delta}$ Thin Films. *Phys. Rev. Lett.* **136**, 066002 (2026).

37 Bhatta, H. C. R. B., Zhang, X., Li, S., Zhong, Y. & Jia, C. Structural and Electronic Evolution of Bilayer Nickelates Under Biaxial Strain. *Commun. Phys.* (2026).

38 Dong, Z. *et al.* Visualization of oxygen vacancies and self-doped ligand holes in $La_3Ni_2O_{7-\delta}$. *Nature* **630**, 847-852 (2024).

39 Liu, Y. B., Mei, J. W., Ye, F., Chen, W. Q. & Yang, F. $s^{\pm}$-Wave Pairing and the Destructive Role of Apical-Oxygen Deficiencies in $La_3Ni_2O_7$ under Pressure. *Phys. Rev. Lett.* **131**, 236002 (2023).

40 Sui, X. *et al.* Electronic properties of the bilayer nickelates $R_3Ni_2O_7$ with oxygen vacancies (R = La or Ce). *Phys. Rev. B* **109**, 205156 (2024).

41 Le, C., Zhan, J., Wu, X. & Hu, J. Opposite-Mirror-Parity Scattering as the Origin of Superconductivity in Strained Bilayer Nickelates. *arXiv:2501.14665* (2025).

42 Sun, J. *et al.* Canonical approach to cation flux calibration in oxide molecular-beam epitaxy. *Phys. Rev. Mater.* **6**, 033802 (2022).

43 Pan, G. A. *et al.* Synthesis and electronic properties of $Nd_{n+1}Ni_nO_{3n+1}$ Ruddlesden-Popper nickelate thin films. *Phys. Rev. Mater.* **6**, 055003 (2022).

44 Lee, J. H. *et al.* Dynamic layer rearrangement during growth of layered oxide films by molecular beam epitaxy. *Nat. Mater.* **13**, 879-883 (2014).

45 Nie, Y. F. *et al.* Atomically precise interfaces from non-stoichiometric deposition. *Nat. Commun.* **5**, 4530 (2014).

46 Sun, Z. P. *et al.* Performance of the BL03U beamline at SSRF. *J. Synchrotron. Radiat.* **27**, 1388-1394 (2020).

47 Blöchl, P. E. Projector augmented-wave method. *Phys. Rev. B* **50**, 17953-17979 (1994).

48 Kresse, G. & Furthmüller, J. Efficient iterative schemes for ab initio total-energy calculations using a plane-wave basis set. *Phys. Rev. B* **54**, 11169-11186 (1996).

49 Perdew, J. P., Burke, K. & Ernzerhof, M. Generalized Gradient Approximation Made Simple. *Phys. Rev. Lett.* **77**, 3865-3868 (1996).